\documentclass[final,nomarks]{dmtcs-episciences}

\usepackage[utf8]{inputenc}
\usepackage{subfigure}

\usepackage{amsmath}
\usepackage{amsthm}
\usepackage{amssymb}
\usepackage{xspace}

\usepackage[round]{natbib}

\newtheorem{fact}{Fact}

\newtheorem{lemma}[fact]{Lemma}
\newtheorem{theorem}[fact]{Theorem}
\newtheorem{proposition}[fact]{Proposition}
\newtheorem{definition}[fact]{Definition}
\newtheorem{corollary}[fact]{Corollary}

\newcommand{\iso}{\ensuremath{\cong}}
\newcommand{\nse}{\ensuremath{\approx}}
\newcommand{\pow}{\ensuremath{\wp}}
\newcommand{\dm}{\ensuremath{diam}}

\newcommand{\ra}{\rightarrow}

\newcommand{\Hom}{\mathrm{Hom}}
\newcommand{\Sur}{\mathrm{Sur}}
\newcommand{\hm}[1]{|\mathrm{Hom}({#1)}|}
\newcommand{\sur}[1]{|\mathrm{Sur}({#1)}|}
\newcommand{\psum}{\mathrm{psum}}

\newcommand{\lequiv}[1]{\equiv^{\mathrm{l}}_{#1}}
\newcommand{\requiv}[1]{\equiv^{\mathrm{r}}_{#1}}

\newcommand{\Ff}{\mathcal{F}}
\newcommand{\Tt}{\mathcal{T}}

\title{Homomorphism Counts to Trees}

\author[Anuj Dawar]{Anuj Dawar\thanks{Research supported in part by UK Research and Innovation (UKRI) under the UK government’s Horizon Europe funding guarantee: grant number EP/X028259/1}}

\affiliation{University of Cambridge, Cambridge, UK.}

\keywords{graph homomorphisms, trees, bipartite graphs}

\begin{document}
  
\publicationdata{vol. 27:3}{2025}{23}{10.46298/dmtcs.13682}{2024-05-30; 2024-05-30; 2024-12-16}{2025-11-06}

\maketitle

\begin{abstract}
  We construct a pair of non-isomorphic, bipartite graphs which are not distinguished by counting the number of homomorphisms to any tree.  This answers a question motivated by Atserias et al. (LICS 2021).  In order to establish the construction, we analyse the equivalence relations induced by counting homomorphisms to trees of diameter two and three and obtain necessary and sufficient conditions for two graphs to be equivalent.  We show that three is the optimal diameter for our construction.
\end{abstract}

\section{Introduction}
For a pair of simple, undirected graphs $G$ and $H$, we write $\Hom(G,H)$ to denote the set of homomorphisms from $G$ to $H$, and $|\Hom(G,H)|$ for the cardinality of this set.
A classic theorem of~\cite{Lovasz67} shows that two finite graphs $G$ and $G'$ are isomorphic if, and only if, for every graph $H$ we have $|\Hom(H,G)| = |\Hom(H,G')|$.  Essentially the same proof can be used to show also that $G$ and $G'$ are isomorphic if, and only if, for every graph $H$ we have $|\Hom(G,H)| = |\Hom(G',H)|$ (for a textbook treatment, see~\cite[Theorem~5.29]{Lovasz12}, and see also~\cite{ChaudhuriV93,Fisk95}).  More generally, any class $\mathcal{F}$ of finite graphs induces two equivalence relations on finite graphs which we call \emph{left-$\mathcal{F}$-equivalence} and \emph{right-$\mathcal{F}$-equivalence} and denote $\lequiv{\Ff}$ and $\requiv{\Ff}$ respectively.  For graphs $G$ and $G'$ we have $G \lequiv{\Ff} G'$ if, and only if, $|\Hom(H,G)| = |\Hom(H,G')|$ for all $H \in \mathcal{F}$ and $G \requiv{\Ff} G'$ if, and only if, $|\Hom(G,H)| = |\Hom(G',H)|$ for all $H \in \mathcal{F}$.  In recent years there has been strong interest in classifying the equivalence relations $\lequiv{\Ff}$ for various classes $\Ff$ as it turns out that suitable choices of $\Ff$ give rise to well-known and much studied equivalence relations.  Lov\'asz' original proof shows that $\lequiv{\Ff}$ is the same as isomorphism when $\Ff$ is the class of connected graphs. Taking $\Ff$ to be the class of graphs of treewidth at most $k$ yields a characterization of the $k$-dimensional Weisfeiler-Leman equivalence, while taking it to be the collection of cycles yields co-spectrality~\cite{Dvorak10,DellGR18}.  For other similar characterizations see~\cite{Grohe20,DawarJR21,Seppelt23,MancinskaR20}.

On the other hand, right-$\Ff$-equivalences for restricted classes $\Ff$ are less well studied.  \cite{AtseriasKW21} initiated a systematic study of these relations and showed that many of the interesting equivalence relations obtained as left-$\Ff$-equivalences are not obtained as right-$\Ff'$-equivalences for any $\Ff'$.
A question arising from this work, communicated by Phokion G.\ Kolaitis, is whether
we can characterize $\requiv{\Tt}$ where $\Tt$ is the class of \emph{finite trees}.  It is clear that, if $T$ is a tree, $\Hom(G,T)$ is non-empty only if $G$ is a bipartite graph and thus the relation $\requiv{\Tt}$ is trivial on graphs which are not bipartite.  On the other hand it was conjectured that for bipartite graphs $\requiv{\Tt}$ is the same as isomorphism.  Here, I give a counterexample to this conjecture.  That is, I describe a construction of a pair of bipartite, non-isomorphic graphs $G$ and $G'$ such that $|\Hom(G,T)| = |\Hom(G',T)|$ for all trees $T$.

Along the way to giving the construction, necessary and sufficient conditions for characterizing the equivalence relations $\requiv{\Tt_1}$, $\requiv{\Tt_2}$ and $\requiv{\Tt_3}$ on connected graphs are considered, where $\Tt_d$ denotes the collection of trees of diameter at most $d$.

\section{Preliminaries}\label{sec:prelim}
I begin with recalling some basic definitions about graphs to fix notation.  All graphs considered are (unless otherwise stated) finite, simple and undirected.  That is, a graph $G=(V,E)$ has a finite set $V$ of vertices and a set $E$ of edges where each edge $e \in E$ is a set of exactly two vertices.  I write $V(G)$ and $E(G)$ for the set of vertices and the set of edges of the graph $G$ respectively.  For two graphs $G$ and $H$, $G \sqcup H$ denotes their \emph{disjoint union}, i.e\ the graph whose vertex set is the disjoint union of $V(G)$ and $V(H)$, and whose edge set is the disjoint union of $E(G)$ and $E(H)$.

For a pair of vertices $u,v$ in a graph $G$, a \emph{walk} from $u$ to $v$ in $G$ is a non-empty sequence $v_0,\ldots, v_p$ of vertices in $V(G)$ with $u = v_0$, $v = v_p$ and such that for each $i \in \{0,\ldots,p-1\}$, $\{v_i,v_{i+1}\} \in E(G)$.  The walk is \emph{closed} if $u = v$.  The walk is a \emph{path} if all vertices in the subsequence $v_1,\ldots, v_{p-1}$ are pairwise distinct and different from $u$ and $v$.  A closed path is a \emph{cycle}.  The \emph{length} of the path $v_0,\ldots, v_p$ is $p$.  The graph $G$ is \emph{connected} if there is a path from $u$ to $v$ for all pairs of distinct vertices $u$ and $v$.  The diameter of $G$, denoted $\dm(G)$ is the least $p$ such that for any pair of distinct $u,v \in V(G)$, there is a path of length at most $p$ from $u$ to $v$.  In particular, $\dm(G) = \infty$ if $G$ is not connected.  

For any graph $G$ and set $S \subseteq
V(G)$ of vertices, $G[S]$ denotes the subgraph of $G$ induced by $S$.  We say that $S$ is an \emph{independent set} in $G$ if $G[S]$ contains no edges.  Also, when $S$ is a set of vertices in a graph $G$, I write $N_G(S)$ to denote the set of neighbours of $S$ in $G$.  When $S = \{s\}$ is a singleton, I also write $N_G(s)$ instead of $N_G(\{s\})$.  Where $G$ is clear from context, I drop the subscript.  The degree of a vertex $v \in V(G)$ is the size of $N(v)$.  Say the graph is $n$-regular if every vertex has degree $n$.

Given two graphs $G$ and $H$, a \emph{homomorphism} $f$ from $G$ to $H$ is a function $f: V(G) \ra V(H)$ such that $f(E(G)) \subseteq E(H)$.  I write $\Hom(G,H)$ to denote the set of all homomorphisms from $G$ to $H$, and $\Sur(G,H)$ to denote the set of \emph{surjective} homomorphisms from $G$ to $H$, i.e.\ a homomorhism $f$ that is surjective onto $V(H)$.  A homomorphism $f$ is an \emph{isomorphism} if it is bijective and $f(E(G)) = E(H)$.  For any collection $\Ff$ of finite graphs, $G \requiv{\Ff} G'$ denotes that $\hm{G,H} = \hm{G',H}$ for all graphs $H \in \Ff$.  I write $G \iso H$ to denote that $G$ and $H$ are isomorphic, i.e.\ there is an isomorphism between them.  In general, we are only interested in graphs up to isomorphism.  Thus, for example, we define \emph{the} complete graph $K_n$ on $n$ vertices below as a concrete graph on vertex set $[n]$ but we also use the term to mean any graph isomorphic to $K_n$.

For any positive integer $n$, $[n]$ denotes the set $\{1,\ldots,n\}$.  For any set $S$, $\pow(S)$ denotes its powerset.  For positive integers $n$ and $m$, $s(n,m)$ is the number of surjections from $[n]$ to $[m]$ (which is non-zero just in case $m \leq n$).  
For any $n$, $S_n$ denotes the \emph{group} of permutations of $[n]$.  Also, for any $n$,  $K_n$ denotes the complete graph on $n$ vertices, i.e.\ $V(K_n) = [n]$ and $E(K_n) = \{\{u,v\} \subseteq [n] \mid u \neq v \}$.  A homomorphism from a graph $G$ to $K_n$ is called an \emph{$n$-colouring} of $G$.  A graph $G$ is said to be \emph{bipartite} if there is a homomorphism from $G$ to $K_2$.  A graph is bipartite if, and only if, it contains no cycle of odd length.  For positive integers $m$ and $n$ with $m \leq n$, $K_{m,n}$ denotes the complete bipartite graph with two parts of $m$ and $n$ vertices respectively.  That is $V(K_{m,n}) = \{0\} \times [m] \cup \{1\} \times [n]$ with $E(K_{m,n})$ the set of all pairs $\{(0,x),(1,y)\}$ with $x \in [m]$ and $y \in [n]$.

A tree is a connected graph containing no cycles.  Since a tree $T$ is bipartite, if there is a homomorphism from $G$ to $T$, then $G$ is also bipartite.  Let $\Tt$ denote the collection of all trees.  For any positive integer $d$, let $\Tt_d$ denote the collection of trees of diameter at most $d$.  

\section{Trees of Diameter $2$}\label{sec:diam2}
In this section, we aim at a classification of bipartite graphs up to equivalence  in $\requiv{\Tt_2}$. There is a relatively clean characterization of this equivalence relation on \emph{connected} bipartite graphs, and this is what we need in subsequent sections.  However, for completeness, I describe the relation for all graphs, and the development also serves as a warm-up for the characterization of $\requiv{\Tt_3}$ on connected graphs that is developed in Section~\ref{sec:diam3}.  

Up to isomorphism, there are two graphs in $\Tt_1$, namely $K_1$ and $K_2$.  The only graphs $G$ for which $\Hom(G,K_1)$ is non-empty are the edgeless graphs, i.e.\ those for which $E(G)$ is empty, and in this case there is exactly one element in $\Hom(G,K_1)$.   Let us now consider homomorphism counts to $K_2$.  
I start by recalling a couple of well known observations.

\begin{fact}\label{fct:disjoint}
  For any graphs $G_1$, $G_2$ and $H$, $\hm{G_1 \sqcup G_2, H} = \hm{G_1,H} \hm{G_2,H}.$
\end{fact}

\begin{fact}\label{fct:k2}
  For any bipartite connected graph $G$, $\hm{G,K_2} = 2$.
\end{fact}

From these two facts it immediately follows that for any bipartite graph $G$, $\hm{G,K_2} = 2^{\gamma(G)}$ where $\gamma(G)$ denotes the number of connected components of $G$.  One consequence is the following.

\begin{proposition}\label{prop:comp}
  For bipartite graphs $G$ and $H$, $\hm{G,K_2} = \hm{H,K_2}$ if, and only if, $G$ and $H$ have the same number of connected components.
\end{proposition}

This then completely characterizes the equivalence relation $\requiv{\Tt_1}$: two bipartite graphs are equivalent if, and only if, they have have the same number of connected components and they are either both edgeless or neither is.  In particular, $\requiv{\Tt_1}$ is isomorphism on edgeless graphs.

To extend the characterization to $\requiv{\Tt_2}$, first observe that this collection of trees contains all graphs $K_{1,p}$ for a positive integer $p$, along with $K_1$ and no others.  In particular, $K_2 \iso K_{1,1,}$.  Thus, the first aim is to characterize the homomorphism counts to graphs $K_{1,p}$.  We do this first for connected graphs.  Let $G$ be a bipartite connected graph with at least two vertices and let $(m,n)$ be the pair of positive integers such that $m \leq n$ and the two parts of $G$ have $m$ and $n$ vertices respectively.  We call the pair $(m,n)$ the \emph{size parameter} of $G$.

\begin{lemma}\label{lem:conn-star}
  For any connected bipartite graph $G$ with size parameter $(m,n)$, $\hm{G,K_{1,p}} = p^m + p^n$.
\end{lemma}
\begin{proof}
  Let $X$ and $Y$ denote the sets of vertices that form the bipartition of $G$, with $|X| = m$ and $|Y| = n$.  We write $s$ and $t_1,\ldots,t_p$ for the vertices of $K_{1,p}$ so that $s$ is a neighbour of $t_i$ for each $i$.

  Since $G$ is connected, there is an even length path between any two distinct vertices in $X$ and between any two distinct vertices in $Y$.  The image of an even length path under a homomorphism is an even length walk and thus
for any homomorphism $h: G \ra K_{1,p}$ we must have that either $h(X) = \{s\}$ and $h(Y) \subseteq \{t_1,\ldots,t_p\}$ or  $h(Y) = \{s\}$ and $h(X) \subseteq \{t_1,\ldots,t_p\}$ .    Now, any function $h: Y \ra \{t_1,\ldots,t_p\}$ can be extended to a homomorphism from $G$ to $K_{1,p}$ by letting $h(x) = s$ for all $x \in X$ and there are exactly $p^n$ such functions.  Similarly,  any function $h: X \ra \{t_1,\ldots,t_p\}$ can be extended to a homomorphism by letting $h(y) = s$ for all $y \in Y$ and there are $p^m$ such functions.  This enumerates all homomorphisms and the result follows.
\end{proof}

Indeed, Fact~\ref{fct:k2} can be obtained as a special case of Lemma~\ref{lem:conn-star} by taking $p=1$.  The following, which we note for further use, is also immediate from the proof of Lemma~\ref{lem:conn-star}.  Recall that $s(n,p)$ denotes the number of surjections from $[n]$ to $[p]$.
\begin{lemma}\label{lem:surj-star}
    For any connected bipartite graph $G$ with size parameter $(m,n)$, $\sur{G,K_{1,p}} = s(m,p) + s(n,p)$.
\end{lemma}

We get the following as consequence of Lemma~\ref{lem:conn-star}, which completely characterizes $\requiv{\Tt_2}$ for connected graphs.
\begin{theorem}\label{thm:diam2}
  The following are equivalent for any pair of connected bipartite graphs $G$ and $H$:
  \begin{enumerate}
  \item $G$ and $H$ have the same size parameter;
  \item $\hm{G,K_{1,2}} = \hm{H,K_{1,2}}$; and 
  \item $G \requiv{\Tt_2} H$.
  \end{enumerate}
\end{theorem}
\begin{proof}
  Clearly (3) implies (2) since $K_{1,2} \in \Tt_2$.  That (2) implies (1) follows from Lemma~\ref{lem:conn-star}, together with the fact that if $m_1,m_2,n_1,n_2$ are positive integers with $m_1 \leq n_1$ and $m_2\leq n_2$, then $2^{m_1}+ 2^{n_1} = 2^{m_2}+2^{n_2}$ if, and only if, $(m_1,n_1) = (m_2,n_2)$.  This is easily verified by writing $2^{m_1}+ 2^{n_1} $ and $2^{m_2}+2^{n_2}$ in base $2$.  Finally, to see that (1) implies (3) note that the only graphs in $\Tt_2$ are $K_{1,p}$ and $K_1$.  The fact that $\hm{G,K_{1,p}} = \hm{H,K_{1,p}}$ follows from Lemma~\ref{lem:conn-star}, and as long as $G$ and $H$ have at least one edge, $\hm{G,K_{1}} = \hm{H,K_{1}} = 0$.
\end{proof}

From Lemma~\ref{lem:conn-star} and Fact~\ref{fct:disjoint} we immediately get the following extension to graphs that are not necessarily connected.
\begin{corollary}\label{cor:bipk}
  Suppose $G$ is a bipartite graph with $k$ connected components which have size parameters $(m_1,n_1),\ldots,(m_k,n_k)$ respectively.  Then
  $$\hm{G,K_{1,p}} = \prod_{i=1}^{k}(p^{m_i}+p^{n_i}).$$
\end{corollary}

It follows that graphs whose components have the same size parameters cannot be distinguished by their right profile to star graphs.
\begin{corollary}\label{cor:suff}
  Suppose $G$ and $H$ are bipartite graphs and there is a bijection $\beta$ between the connected  components of $G$ and those of $H$ such that each component $C$ of $G$ has the same size parameter as $\beta(C)$.  Then, $G \requiv{\Tt_2} H$.
\end{corollary}
This gives a sufficient, but not necessary, condition for two graphs to be indistinguishable by their homomorphism counts to trees in $\Tt_2$.  Nonetheless, it already establishes that $\requiv{\Tt_2}$ is a rather weak discriminator.  It cannot determine more than the size parameters of the components of $G$.

To give a complete characterization, I first introduce some notation.  Let $\vec{d} = (d_1,\ldots,d_k)$ be a $k$-tuple of non-negative integers.  Write $\psum(\vec{d})$ for the \emph{multiset of partial sums} of $\vec{d}$.  That is the multiset $\{\!\!\{ \sum_{i \in S} d_ i \mid S \subseteq \{1,\ldots,k\} \}\!\!\}$.

Let $G$ be a bipartite graph with $\gamma(G) = k$ (i.e.\ $G$ has $k$ components).  Let $(m^G_1,n^G_1),\ldots,(m^G_k,n^G_k)$ be the size parameters of the components of $G$.  Write $\vec{d}^G$ for the $k$-tuple $(d^G_1,\ldots,d^G_k)$ where $d^G_i = n^G_i - m^G_i$ (recall that by definition of size parameters $m^G_i \leq n^G_i$). 

\begin{theorem}\label{thm:T2}
 For any bipartite graphs $G$ and $H$, $G \requiv{\Tt_2} H$ if, and only if, $\gamma(G) = \gamma(H)$,  $\sum_{i=1}^{\gamma(G)} m^G_i = \sum_{i=1}^{\gamma(H)} m^H_i$ and $\psum(\vec{d}^G) = \psum(\vec{d}^H)$.
\end{theorem}
\begin{proof}
  For the direction from right to left let $k= \gamma(G) = \gamma(H)$ and note that, assuming $G$ and $H$ are not edgeless, for any  $p$ we have
    \begin{eqnarray*}
    \hm{G,K_{1,p}} & = & \prod_{i=1}^{k}(p^{m^G_i}+p^{n^G_i}) \\
               & = & \sum_{S \subseteq \{1,\ldots,k\}} \prod_{i \in S}p^{m^G_i}\prod_{j \not\in S}p^{n^G_j} \\
               & = & \sum_{S \subseteq \{1,\ldots,k\}} \prod_{i \in S}p^{m^G_i}\prod_{j \not\in S}p^{m^G_j+d^G_j} \\
               & = & \sum_{S \subseteq \{1,\ldots,k\}} \prod_{i =1}^kp^{m^G_i}\prod_{j \not\in S}p^{d^G_j} \\
               & = & p^{\sum_{i =1}^k m^G_i}(\sum_{S \subseteq \{1,\ldots,k\}}p^{\sum_{j \in S} d^G_j}) \\
               & = &  p^{\sum_{i =1}^k m^G_i}(\sum_{\delta \in \psum(\vec{d}^G)}p^{\delta}) \\
               & = & p^{\sum_{i =1}^k m^H_i}(\sum_{\delta \in \psum(\vec{d}^H)}p^{\delta}) \\
               & = & \hm{H,K_{1,p}},
    \end{eqnarray*}
    where the first equation is by Corollary~\ref{cor:bipk}.
    
    For the other direction, since $G \requiv{\Tt_2} H$, we have $\hm{G,K_2} = \hm{H,K_2}$ and hence $\gamma(G) = \gamma(H)$ by Proposition~\ref{prop:comp}.  Letting $k = \gamma(G) = \gamma(H)$, 
    consider again the expression obtained above, namely
     \begin{equation}\label{hmG}
    \hm{G,K_{1,p}} = \sum_{S \subseteq \{1,\ldots,k\}} p^{\sum_{i =1}^k m^G_i + \sum_{j\in S} d^G_j}.
    \end{equation}
     This is a sum of $2^k$ terms each of which is a power of $p$ and so can be written as a polynomial in $p$ where each coefficient is at most $2^k$.  Thus, since
  \begin{equation}
    \label{hmH}
  \hm{H,K_{1,p}} = \sum_{S \subseteq \{1,\ldots,k\}} p^{\sum_{i =1}^k m^H_i + \sum_{j\in S} d^H_j}
  \end{equation}
  is a similar sum, by taking $p > 2^k$, the assumption that $\hm{G,K_{1,p}} = \hm{H,K_{1,p}}$ implies that the coefficients of the corresponding powers of $p$ are equal in the two polynomials.

      The smallest power of $p$ occurring on the right hand side of~(\ref{hmG}) is $\sum_{i =1}^k m^G_i$ and on the right hand side of~(\ref{hmH}) is $\sum_{i =1}^k m^H_i$ and so we immediately get $\sum_{i =1}^k m^G_i = \sum_{i =1}^k m^H_i.$

  Similarly, for any $\delta \in \psum(\vec{d}^G)$, the coefficient of $p^{\sum_{i =1}^k m^G_i + \delta}$ on the right hand side of~(\ref{hmG}) is exactly the multiplicity of $\delta$ in $\psum(\vec{d}^G)$ and so we have that $\delta$ must have the same multiplicity in  $\psum(\vec{d}^H)$.  Thus, $\psum(\vec{d}^G) = \psum(\vec{d}^H)$ as multisets.
\end{proof}

It is an interesting question whether the requirement $\psum(\vec{d}^G) = \psum(\vec{d}^H)$ in Theorem~\ref{thm:T2} can be replaced by the apparently stronger statement that $\vec{d}^G$ and $\vec{d}^H$ are the same as multisets.

\section{Trees of Diameter $3$}\label{sec:diam3}

We now consider the class $\Tt_3$ of trees of diameter $3$ and establish a sufficient condition for connected graphs $G$ and $H$ to have $G \requiv{\Tt_3} H$.  This is used in the next section to prove the required properties of our main construction.

Note that $\Tt_3$ contains all trees in $\Tt_2$ and in addition all trees of diameter exacty $3$.  Each such tree contains
exactly two vertices $u$ and $v$ of degree greater than $1$.  These
two vertices are neighbours and every other vertex is either a
neighbour of $u$ or a neighbour of $v$.  We write $B_{p,q}$ for such a
graph where $u$ has $p$ neighbours other than $v$ and $v$ has $q$
neighbours other than $u$.  We can think of $B_{p,q}$ as the graph formed by taking disjoint copies of $K_{1,p}$ and $K_{1,q}$ and putting an edge between the two singleton parts.  Thus, to characterise the relation $\requiv{\Tt_3}$, we need to study the sizes of $\Hom(G,B_{p,q})$ for bipartite graphs $G$.

We begin by defining an equivalence relation on bipartite graphs.  Recall that for a connected bipartite graph $G$, the \emph{size parameter} of $G$ is the pair $(m,n)$ of positive integers such that $m \leq n$ and the two parts of the bipartition of $G$ have $m$ and $n$ vertices respectively. 

\begin{definition}
  Let $G$ and $H$ be two connected bipartite graphs and let $X^G, Y^G$ and $X^H,Y^H$ be the bipartitions of $G$ and $H$ respectively, with $|X^G| \leq |Y^G|$ and $|X^H| \leq |Y^H|$.  We say that $G$ and $H$ are \emph{neighbourhood size equivalent} and write $G \nse H$ if they have the same size parameter (i.e.\ $|X^G| = |X^H|$ and $|Y^G| = |Y^H|$) and there is a bijection $\eta: \pow (X^G) \rightarrow \pow(X^H)$ such that for all $S \subseteq X^G$:
  \begin{enumerate}
  \item $|S| = |\eta(S)|$; and 
  \item $|N(S)| = |N(\eta(S))|$.
  \end{enumerate}
\end{definition}
The bijection $\eta$ in this definition is required to preserve the sizes of sets and the sizes of neighbourhoods.  
In other words, $G \nse H$ if, and only if, they have size parameter
$(m,n)$ and for each $p \leq m$, the multiset of integers obtained by
taking the sizes $|N(S)|$ of neighbourhoods of subsets $S$ of $X$ of
size $p$ is the same for $G$ and $H$.

I now show that the equivalence relation $G \nse H$ is sufficient to ensure that $G \requiv{\Tt_3} H$.

\begin{lemma}\label{lem:bistar}
  For any pair of bipartite connected graphs $G$ and $H$ and any positive
  integers $p$ and $q$, if $G \nse H$ then $\hm{G,B_{p,q}} =
  \hm{H,B_{p,q}}$ and $\sur{G,B_{p,q}} = \sur{H,B_{p,q}}$.
\end{lemma}
\begin{proof}
  Let $(m,n)$ be the size parameter of $G$ and let $(X^G,Y^G)$ be the bipartition of $V(G)$ such that $|X^G|= m$ and $|Y^G| = n$.  Let $u$ and $v$ be the non-leaf vertices of $B_{p,q}$ with $S$ the set of $p$ neighbours of $u$ other than $v$ and $T$ the set of $q$ neighbours of $v$ other than $u$.

Let $h: G \ra B_{p,q}$ be a homomorphism.  Then, $h^{-1}(S)$, if it is non-empty, is either a subset of $X^G$ or of $Y^G$.  If $\emptyset \neq h^{-1}(S) \subseteq X^G$, then $h^{-1}(T) \subseteq  Y^G$.  Indeed, $h^{-1}(T) \subseteq Y^G \setminus N(h^{-1}(S))$, since all neighbours of vertices in $h^{-1}(S)$ must map to $u$.  But, for a fixed set $A \subseteq X^G$, any function mapping $A$ to $S$ and  $B = (Y^G \setminus N(A))$ to $T \cup \{u\}$ extends uniquely to a homomorphism that maps $X^G \setminus A$ to $v$ and $N(A)$ to $u$.  Similarly, if $h^{-1}(T) \subseteq X^G$, the situation is mirrored.  Putting this all together, we get
\begin{equation}\label{eqn:first}
  \hm{G,B_{p,q}} = \sum_{A \subseteq X^G} p^{|A|}((q+1)^{|Y^G \setminus N(A)|}) +  \sum_{A \subseteq X^G} q^{|A|}((p+1)^{|Y^G \setminus N(A)|}).
 \end{equation}

 Now, let $X^H$ and $Y^H$ be the bipartition of $V(H)$ such that $|X^H| = m $ and $|Y^H| = n $ and  $\eta : \pow(X^G) \ra \pow(X^H)$ is the bijection witnessing that $G \nse H$.  Then, we have:
\begin{eqnarray*}
  \hm{G,B_{p,q}} & = & \sum_{A \subseteq X^G} p^{|A|}((q+1)^{|Y^G \setminus N(A)|}) +  \sum_{A \subseteq X^G} q^{|A|}((p+1)^{|Y^G \setminus N(A)|}) \\
                  & = & \sum_{A \subseteq X^G} p^{|\eta(A)|}((q+1)^{|Y^H \setminus N(\eta(A))|}) +  \sum_{A \subseteq X^G} q^{|\eta(A)|}((p+1)^{|Y^H \setminus N(\eta(A))|}) \\
                  & = & \sum_{A \subseteq X^H} p^{|A|}((q+1)^{|Y^H \setminus N(A)|}) +  \sum_{A \subseteq X^H} q^{|A|}((p+1)^{|Y^H \setminus N(A)|}) \\
  & =& \hm{H,B_{p,q}},
\end{eqnarray*}
where the first line is from~(\ref{eqn:first}), the second follows from the fact that $|A| = | \eta(A)|$ and $|N(A)| = |\eta(N(A))|$ for all $A \subseteq X^G$, the third follows from the fact that $\eta$ is a bijection so summing over  $\eta(A)$ for all subsets $A$ of $X^G$  is the same as summing over all subsets $A$ of $X^H$.

Counting surjective homomorphisms introduces a slight complication.  We can choose a set $A$ to be $h^{-1}(S)$ and count the surjective maps to $S$, but we must have $A \neq X^G$ so that at least one element maps to $v$.   Thus, we get:
\begin{eqnarray}\label{eqn:surj}
  \sur{G,B_{p,q}} & = & \sum_{A \subsetneq X^G} s(|A|,p)(s(|Y^G \setminus N(A)|,q) + s(|Y^G \setminus N(A)|,q+1)) + \\
  & \mbox{} & \mbox{} + \sum_{A \subsetneq X^G} s(|A|,q)(s(|Y^G \setminus N(A)|,p)+s(|Y^G \setminus N(A)|,p+1)). 
\end{eqnarray}
To explain this sum, in the first summation, we sum over all sets $A$ that might appear as $h^{-1}(S)$.  For each of these, there are $s(|A|,p)$ surjective maps to $S$ (this number might be zero).  For each such map, we count the ways in which $Y^G \setminus N(A)$ maps surjectively to either $T$ or $T \cup \{u\}$.  Note that either of these yields a surjective map from $Y^G$ to $T \cup \{u\}$ since $N(A)$ must be non-empty when $A$ is, and $N(A)$ necessarily maps to $u$.

Just as with the expression for $\hm{G,B_{p,q}}$ given in~(\ref{eqn:first}), we can see that the expression given in~(\ref{eqn:surj}) for $\sur{G,B_{p,q}}$ is completely determined by the sizes of sets $A$ and $N(A)$.  Thus the bijection $\eta$ guarantees that $\sur{G,B_{p,q}} = \sur{H,B_{p,q}}$.
\end{proof}

For bipartite, connected graphs $G$ and $H$, the equivalence relation $G \nse H$ guarantees in particular that $G$ and $H$ have the same size parameter and thus, by Lemma~\ref{lem:conn-star} that $G \requiv{\Tt_2} H$.  Together with Lemma~\ref{lem:bistar} this implies the following.
\begin{theorem}\label{thm:nse}
  For bipartite, connected graphs $G$ and $H$, $G \nse H$ implies $G \requiv{\Tt_3} H$. 
\end{theorem}

\paragraph*{Remark.}
In an earlier version of this paper, I conjectured that the converse of Theorem~\ref{thm:nse} holds.  It was subsequently pointed out to me by Dan Kr\'{a}l that this is not the case.  A counter-example was constructed by 
Igor Balla, Jan B\"{o}ker, Daniel Ilkovič, Bartek Kielak and Dan Kr\'{a}l  and I include it here with their permission.

Let $X = \{a,b,c\}$ and $Y = \{d,e,f,g,h\}$ be two sets of vertices.  Let $H_1$ and $H_2$ be two graphs with bipartition $X$ and $Y$ where the edges of $H_1$ are all pairs in $X \times Y$ \emph{except} $(a,d), (a,e), (a,f), (b,g)$ and $(b,h)$ while the edges of $H_2$ are all pairs in $X \times Y$ \emph{except} $(a,d), (a,e), (a,f), (b,g)$ and $(c,g)$.  Thus, each of $H_1$ and $H_2$ has ten edges.  We have $H_1 \not\nse H_2$ since the degrees of the three vertices in $X$ are different in the two graphs.  In $H_1$, the degrees are $2, 3$ and $5$, while in $H_2$ they are $2, 4$ and $4$.  To see that $H_1 \requiv{\Tt_3} H_2$, we use another characterization of $\hm{G,B_{p,q}}$ for a bipartite graph $G$.  Thus, let $G$ be any bipartite connected graph, with a bipartition into sets $X$ and $Y$ and let $\mathcal{I}$ denote the collection of all independent sets of $G$.  Then
\begin{equation}\label{eqn:kral}
  \hm{G, B_{p,q}} = \sum_{A \in \mathcal{I}} \left( p^{|A \cap X|}q^{|A\cap Y|} + p^{|A \cap Y|}q^{|A\cap X|} \right) .
\end{equation}    
Now, for any independent set $A$, say its \emph{type} is $(m,n)$ if $|A\cap X|= m$ and $|A \cap Y| = n$.  Then, for all $(m,n)$ with $0 \leq m \leq 3$ and $0 \leq n \leq 5$ except for $(1,2)$ and $(2,1)$, it can be verified by inspection that $H_1$ and $H_2$ have the same number of independent sets of type $(m,n)$.  Also $H_1$ has four independent sets of type $(1,2)$ and none of type $(2,1)$ and $H_2$ has three independent sets of type $(1,2)$ and one of type $(2,1)$.  Thus, by~(\ref{eqn:kral}), we have $\hm{H_1, B_{p,q}} = \hm{H_2, B_{p,q}}$.

\section{Counterexample}\label{sec:counter}

With the results of the previous two sections in hand, we are ready to construct our example of a pair of non-isomorphic, bipartite graphs $G$ and $H$ for which $G \requiv{\Tt} H$.  Here $\Tt$ is the class of \emph{all} trees, but it suffices to consider trees with diameter at most $3$ if we construct $G$ and $H$ to have diameter $3$.  This argument is formalized in the following theorem.
\begin{theorem}\label{thm:homtree}
  If $G$ and $H$ are connected bipartite graphs of diameter at most $3$, and $G \nse H$,
  then $G \requiv{\Tt} H$.
\end{theorem}
\begin{proof}
    For any graph $G$ of diameter $3$ and any tree $T$, the image of $G$
  under a homomorphism to $T$ must be a subtree of $T$ of diameter at
  most $3$.  Thus, we have
  $$ \hm{G,T} = \sum_{T'} \sur{G,T'} $$
  where the sum is taken over all induced subtrees $T'$ of $T$ of diameter at
  most $3$.

  Since $G$ and $H$ are both connected, they are either both $K_1$ and hence isomorphic, or they both contain an edge and so $\sur{G,K_1} = \sur{H,K_1} = 0$ and $\sur{G,K_2} = \sur{H,K_2} = 2$.  The fact that $G \nse H$ ensures that they have the same size parameter and hence $\sur{G,K_{1,p}} = \sur{H,K_{1,p}}$ for all $p$ by Lemma~\ref{lem:surj-star}.  Furthermore  $G \nse H$ implies $\sur{G,B_{p,q}} = \sur{H,B_{p,q}}$ by Lemma~\ref{lem:bistar}.  Thus   $\sur{G,T'} = \sur{H,T'}$ for all trees $T'$ of diameter at most $3$.
\end{proof}

The aim now is to construct the pair of graphs $G$ and $H$ which both have diameter $3$ and for which $G \nse H$ but $G \not\iso H$.  I give the construction of the graphs in stages, proving relevant properties along
the way.  For this, it is useful to give a rather more general construction of a family of graphs, which includes the specific examples we are interested in.

Let $n \geq 3$ be an integer and fix four disjoint sets of $n$ vertices each: $A = \{a_1,\ldots,a_n\},
B= \{b_1,\ldots,b_n\}, C = \{c_1,\ldots,c_n\}$ and $D = \{d_1,\ldots,d_n\}$.

Let $G^-$ denote the graph on the set of vertices $A\cup
B\cup C \cup D$ whose edge set is the union of the following three
sets:
\begin{itemize}
\item $E_{AB}  = \{\{a_i,b_j\} \mid i \neq j \}$; 
\item $E_{BC} =  \{\{b_i,c_i\} \mid 1 \leq i \leq n  \}$; 
\item $E_{CD} =  \{\{c_i,d_j\} \mid i \neq j \}$; 
\end{itemize}
In other words, each of  $G^-[A \cup B]$ and $G^-[C \cup D]$ is an
$(n-1)$-regular bipartite graph on two sets of $n$ vertices, and
$G^-[B \cup C]$ is a perfect matching.  Note that there is a graph
$G^-$ for each value of $n$, but we elide mention of $n$ when it is
clear from the context.

Now, for each permutation $\pi \in S_n$, define $G^-_{\pi}$ to be the graph on the vertex set $A \cup B \cup C \cup D$ whose edges are $E_{AB} \cup E_{BC} \cup E_{CD}
 \cup E_{AD}$ where
 \begin{quote}
   $E_{AD} =  \{\{a_i,d_{\pi(i)}\} \mid 1 \leq i \leq n  \}$; 
 \end{quote}
 so that $G^-_{\pi}[A\cup D]$ is also a perfect matching.

It is clear that each $G^{-}_{\pi}$ is an $n$-regular bipartite connected graph with a bipartition into sets $A \cup C$ and $B \cup D$, where each part has $2n$ vertices.

Out next aim is to classify some isomorphisms between these graphs.  We are particularly interested in isomorphisms that fix the sets $A, B, C$ and $D$ and for ease of reference, I give them a name.  Say a map $\alpha: A\cup B \cup C \cup D \rightarrow A\cup B \cup C \cup D$ is \emph{clean} if $\alpha(A) = A$, $\alpha(B) = B$, $\alpha(C) = C$ and $\alpha(D) = D$.

It is useful to recall some basic group-theoretic notions.  Recall that for any group $\mathcal{G}$, the \emph{order} of $x \in \mathcal{G}$ is the least positive integer $i$ such that $x^i$ is the identity element.  Also, recall that 
two elements $x, y \in \mathcal{G}$ are said to be \emph{conjugate} if there is some $z \in \mathcal{G}$ such that $x = zyz^{-1}$.  Conjugacy is an equivalence relation and, in particular, it preserves the order of elements.  That is to say that two elements that are conjugate must have the same order. 

\begin{lemma}\label{lem:conjugate}
 For $\pi_1,\pi_2 \in S_n$, there is a clean isomorphism between  $G^{-}_{\pi_1}$ and $G^{-}_{\pi_2}$ if, and only if, $\pi_1$ and $\pi_2$ are conjugate.
\end{lemma}
\begin{proof}
  Note first of all that any clean isomorphism $\alpha$ between  $G^{-}_{\pi_1}$ and $G^{-}_{\pi_2}$ is also an automorphism of $G^{-}$.  I claim that if $\alpha$ is a clean automorphism of $G^-$ then it is given by some permutation $\sigma \in S_n$ so that
  $\alpha(x_i) = x_{\sigma(i)}$ for $x \in \{a,b,c,d\}$ and $i \in  \{1,\ldots,n\}$.   Moreover, any $\sigma \in S_n$ gives rise to an automorphism of $G^{-}$ in this way.  To see this, first note that since $\alpha(A) = A$ by
  assumption, there must be a $\sigma \in S_n$ such that $\alpha(a_i)
  = a_{\sigma(i)}$ for all $i \in \{1,\ldots,n\}$.   Now, since $b_i$ is the
  only vertex in $B$ that is not a neighbour of $a_i$, we must have
  $\alpha(b_i) = b_{\sigma(i)}$.  Similarly, as $c_i$ is the only
  neighbour of $b_i$ in $C$, we get $\alpha(c_i) = c_{\sigma(i)}$ and
  finally as $d_i$ is the only non-neighbour of $c_i$ in $D$, we must
  have $\alpha(d_i) = d_{\sigma(i)}$.  

  For $\alpha$ to be an isomorphism between $G^{-}_{\pi_1}$ and $G^{-}_{\pi_2}$, it must further preserve the edges between $A$ and $D$.  There is an edge between $a_i$ and $d_j$ in $G^{-}_{\pi_1}$ if, and only if, $j = \pi_1(i)$.  So, $\alpha$ is an isomorphism if, and only if,  for all $i,j$:
  $$j  =\pi_1(i) \quad \mbox{ if, and only if, } \quad \sigma(j) = \pi_2(\sigma(i)).$$
But this is equivalent to $\sigma\pi_1\sigma^{-1} = \pi_2$.
\end{proof}

We next show that the equivalence relation $\nse$ is weaker via the following lemma.  For this, recall that a fixed point of a permutation $\pi
\in S_n$ is a value $i \in \{1,\ldots,n\}$ such that $\pi(i) = i$.

\begin{lemma}\label{lem:nse}
  If $\pi_1,\pi_2 \in S_n$ are two permutations with the same number
  of fixed points, then $G^-_{\pi_1} \nse G^-_{\pi_2}$.
\end{lemma}
\begin{proof}
  The two graphs both have size parameter $(2n,2n)$, and I consider
  all subsets $S$ of $A \cup C$ through a series of mutually exclusive
  cases and define
  the neighbourhood-size preserving bijection $\eta$ on them.
  Clearly, for $S = \emptyset$, $\eta(S) = \emptyset$ is the only size
  preserving choice, so we consider non-empty sets.

  \noindent
  \emph{Case 1:} $S$ is a singleton.  Then $|N(S)| = n$ since
  $G^-_{\pi}$ is always an $n$-regular graph.  Any bijection between
  $A \cup C$ and itself will do and for concreteness, let $\eta(S) =
  S$.

  \noindent
  \emph{Case 2:} $S \subseteq A$ and $|S| \geq 2$.  In this case,
  necessarily,  $B \subseteq N(S)$ and $S$ has exactly $|S|$
  neighbours in $D$.  Thus, $|N(S)| = n + |S|$ in either graph.
  Letting $\eta(S) = S$ satisfies the requirements.

   \noindent
  \emph{Case 3:} $S \subseteq C$ and $|S| \geq 2$.  This is entirely
  analogous to the previous case.

   \noindent
  \emph{Case 4:} $|S \cap A| \geq 2$ and $|S \cap C| \geq 2$.  In this
  case $B \cup D \subseteq N(S)$ and so necessarily $|N(S)| = 2n$ in
  both graphs.  Thus, taking $\eta(S) = S$ satisfies the requirements.

  \noindent
  \emph{Case 5:} $|S \cap C| \geq 2$ and $|S \cap A| = 1$.  I claim
  that letting $\eta(S) = S$ also satisfies the size requirements in
  this case.  Let $a_i$ be the unique element of $S \cap A$.  Since $S
  \cap C$ has at least two elements, $D \subseteq N(S)$.  Moreover,
  since every element of $B$ apart from $b_i$ is a neighbour of $a_i$,
  we have that $|N(S)| = 2n$ if $c_i \in S$ and $|N(S)| = 2n -1$
  otherwise.  This is true in either graph, and so $|N(S)| = |N(\eta(S))|$.
  
  \noindent
  \emph{Case 6:} $|S \cap A| \geq 2$ and $|S \cap C| = 1$.  This is
  the first case where the identity map will not suffice.  We need to
  apply a permutation to the set $C$.  Otherwise, it is similar to the
  previous case.
  
  Let $Z = S \cap A$ and let $c_i$ be the unique element of
  $S \cap C$.  Let $\eta(S) = Z \cup \{c_{\pi_2\pi_1^{-1}(i)}\}$.
  Clearly this is a size-preserving bijection on the collection of sets satisfying
  $|S \cap A| \geq 2$ and $|S \cap C| = 1$.  To see that this
  satisfies the neighbourhood size condition, note that because
  $|S\cap A| \geq 2$, we have $B \subseteq N(S)$.  Moreover, since the
  neighbours of $c_i$ in $D$ are all elements except $d_i$, we have
  that $|N(S)| = 2n$ if $d_i \in N(Z)$ and $|N(S)| = 2n-1$ otherwise.
  The former case happens in $G^-_{\pi_1}$ precisely when
  $a_{\pi^{-1}(i)} \in Z$.  But this ensures that
  $d_{\pi_2\pi_1^{-1}(i)} \in N_{G^-_{\pi_2}}(Z)$ and so
  $|N_{G^-_{\pi_2}}(\eta(S))| = 2n$.

  \noindent
  \emph{Case 7:}  $|S \cap A| = 1$ and $|S \cap C| = 1$.  This is the
  final case and the only one where we need the assumption that
  $\pi_1$ and $\pi_2$ have the same number of fixed points.

  Let $S = \{a_i,c_j\}$.  There are $n^2$ such sets, and we divide
  them into four cases. Note that every element in $B$ apart from
  $b_i$ is a neighbour of $a_i$ and every element in $D$ apart from
  $d_j$ is a neighbour of $c_j$.  Thus, $N(S)$ contains at least
  $2n-2$ elements.  Whether it has $2n-2$, $2n-1$ or $2n$ depends on
  whether $b_i$ and $d_j$ are in it.  This gives the four cases below.  We deal with both graphs at the same time so, in the following,
  $\pi$ refers to either of the permutations $\pi_1$ or $\pi_2$.
  \begin{description}
  \item[(a)] $i \neq j$ and $\pi(i) \neq j$.  This means $b_i$ is
    not a neighbour of $c_j$ and $d_j$ is not a neighbour of $a_i$.
    Thus $|N(S)| = 2n-2$.
  \item[(b)] $i \neq j$ and $\pi(i) = j$.  In this case, $b_i$ is not
    a neighbour of $c_j$ but $d_j$ is a neighbour of $a_i$, so $|N(S)|
    = 2n-1$.
  \item[(c)] $i=j$ and $\pi(i) \neq j$.  Here, $b_i$ is a neighbour of
    $c_j = c_i$ but $d_j$ is not a neighbour of $a_i$, so $|N(S)|
    = 2n-1$.
  \item[(d)] $i = j$ and $\pi(i) = j$.  Finally, in this case $b_i$ is
    a neighour of $c_j$ and $d_j$ is a neighbour of $a_i$ so $|N(S)| = 2n$.
  \end{description}

  Let $p$ denote the number of fixed-points of $\pi$.  Then, there are
  exactly $p$ pairs $(i,j)$ satisfying case~\textbf{(d)} above and
  therefore exactly $n-p$ pairs satisfying each of the
  cases~\textbf{(b)} and~\textbf{(c)}, leaving $n^2-2n+p$ pairs
  satisfying case~\textbf{(a)}.  Hence, as long as $\pi_1$ and $\pi_2$
  have the same number of fixed points, there is a bijection on the
  sets of pairs that guarantees $|N_{G^-_{\pi_1}}(S)| =
  |N_{G^-_{\pi_2}}(\eta(S))|$. 
  
\end{proof}

Note that the bijection $\eta$ constructed in the proof of
Lemma~\ref{lem:nse} has the property that, not only $|\eta(S)| = |S|$,
but $|\eta(S) \cap A| = |S \cap A|$ and $|\eta(S) \cap C| = |S \cap C|$ which is
important for the next step of the construction.

For any $n \geq 3$ and $\pi \in S_n$ define the graph $G_{\pi}$ to be
the graph obtained from $G^-_{\pi}$ by adding three additional
vertices $e,f,g$ so that:
\begin{enumerate}
\item $N(e) = B \cup D \cup \{f,g\}$; 
\item $N(f) = A \cup C \cup \{e\}$; and
\item $N(g) = C \cup \{e\}$.
\end{enumerate}

We now observe some useful facts about the graphs $G_{\pi}$.
First, note that  $G_{\pi}$ is bipartite.  The bipartition is given by the sets $X
  = A \cup C \cup \{e\}$ and $Y = B \cup D \cup \{f,g\}$.  Secondly,
  $G_{\pi}$ has diameter at most $3$.  This is because the vertices $e$ and
  $f$ are adjacent to all vertices in the other part.  Indeed, this is the purpose of these vertices, to force the diameter to be at most $3$ while preserving other relevant properties.  The purpose of the vertex $g$ is to distinguish the sets $A$, $B$, $C$ and $D$ from each other, forcing all isomorphisms to be clean.

  \begin{lemma}\label{lem:clean}
    For $\pi_1,\pi_2 \in S_n$, $G_{\pi_1} \iso G_{\pi_2}$ if, and only
    if, there is a clean isomorphism between $G^-_{\pi_1}$ and $G^-_{\pi_2}$.
  \end{lemma}
  \begin{proof}
    First, suppose $\alpha$ is a clean isomorphism between
    $G^-_{\pi_1}$ and $G^-_{\pi_2}$.  We can extend it to an
    isomorphism between $G_{\pi_1}$ and $G_{\pi_2}$ by letting  $\alpha(e) = e$, $\alpha(f) = f$ and $\alpha(g) = g$.  Since $\alpha$ fixes $A$,
    $B$, $C$ and $D$ setwise  by assumption, it also fixes the
    neighbourhoods of each of $e$, $f$ and $g$ and so is an
    isomorphism.

    In the other direction, suppose $\alpha$ is an isomorphism from
    $G_{\pi_1}$ to $G_{\pi_2}$.  I argue that its restriction to the
    vertex set of $G^-$ is a clean isomorphism between $G^-_{\pi_1}$ and
    $G^-_{\pi_2}$.   Since
    $G_{\pi_1}$ and $G_{\pi_2}$ are connected and bipartite, $\alpha$
    must preserve the bipartition and since $|X| \neq |Y|$ we have
    that $\alpha(X) = X$ and $\alpha(Y) = Y$.  Since $e$ is the only
   vertex of degree $2n+2$ and $f$ the only vertex of degree $2n+1$,
   we have $\alpha(e) = e$ and $\alpha(f) = f$.  The vertices in $C$
   are the only vertices in $X$ of degree $n+2$ and thus $\alpha(C) =
   C$.  Since $g$ is the only vertex, apart from $f$ that is a
   neighbour to all vertices in $C$, we have $\alpha(g) = g$.
   Then the vertices in $B$ are the only ones with exactly one
   neighbour in $C$ and those in $D$ the only ones with exactly $n-1$
   neighbours in $C$, and therefore $\alpha(B) = B$ and $\alpha(D) =
   D$.  What is left is $A$, so $\alpha(A) = A$.
 \end{proof}

 The following is now immediate from Lemmas~\ref{lem:conjugate} and~\ref{lem:clean}.
 \begin{corollary}\label{cor:iso}
For any $\pi_1,\pi_2 \in S_n$,    $G_{\pi_1} \iso G_{\pi_2}$ if, and only    if, $\pi_1$ and $\pi_2$  are conjugate.
 \end{corollary}

 Now, we show that adding the extra vertices $e$, $f$ and $g$ does not break the equivalence relation $\nse$.

  \begin{lemma}\label{lem:fullnse}
    For any $\pi_1,\pi_2 \in S_n$,  if $G^-_{\pi_1} \nse G^-_{\pi_2}$, then $G_{\pi_1} \nse G_{\pi_2}$.
  \end{lemma}
  \begin{proof}
  Both graphs $G_{\pi_1}$ and $G_{\pi_2}$ have size parameter
  $(2n+1,2n+2)$.    Let $\eta: \pow(A \cup C) \ra \pow(A \cup C)$ be
  the bijection witnessing that   $G^-_{\pi_1} \nse G^-_{\pi_2}$.   We
  extend it to a bijection $\eta': \pow(X) \ra \pow(X)$ by having, 
  for all $S \subseteq A \cup C$:   $\eta'(S) = \eta(S)$ and $\eta'(S
  \cup\{e\}) = \eta(S) \cup \{e\}$.  This bijection is clearly size
  preserving.  To see that it preserves sizes of neighbourhoods, note that:
  \begin{enumerate}
  \item $|N_{G_\pi}(S)| = 2n+2$ if $e \in S$;
  \item $|N_{G_\pi}(S)| = |N_{G^-_{\pi}}(S)| + 1$ if $S \subseteq A$; and 
  \item $|N_{G_\pi}(S)| = |N_{G^-_{\pi}}(S)| + 2$ if $e \not\in S$ and $S
      \cap C \neq \emptyset$.
  \end{enumerate}
  Since $\eta$, as defined in the proof of Lemma~\ref{lem:nse} has the
  property that $S \cap C = \emptyset$ if, and only if, $\eta(S) \cap
  C = \emptyset$, it follows that for all $S$, $|N(\eta'(S)| = |N(S)|$.
\end{proof}

From Lemma~\ref{lem:nse} and Lemma~\ref{lem:fullnse}, we can now conclude the following corollary.
\begin{corollary}\label{cor:nse}
  If $\pi_1$ and $\pi_2$ have the same number of fixed-points, then
  $G_{\pi_1} \nse G_{\pi_2}$. 
\end{corollary}

It follows from Corollary~\ref{cor:nse} and
Corollary~\ref{cor:iso} that if $\pi_1$ and $\pi_2$ are permutations
that are not conjugate in $S_n$ but they have the same number of fixed
points, then $G_{\pi_1} \not\iso G_{\pi_2}$ but $G_{\pi_1} \nse
G_{\pi_2}$.  For any $n \geq 4$, there exist such pairs of
permutations. In particular, fix $n = 4$ and let $\pi_1 = (12)(34)$
and let $\pi_2 = (1234)$.  These are not conjugate, as $\pi_1$ has
order $2$ and $\pi_2$ has order $4$, but neither of them has any
fixed-points.  The two graphs $G= G_{(12)(34)}$ and $H = G_{(1234)}$ are the
pair of graphs which witness the main claim of this paper.  That is, $G \not\iso H$ and $G \requiv{\Tt} H$.

\section{Conclusion}
The main construction in the present paper shows that the equivalence relation $\requiv{\Tt}$ is not the same as isomorphism on bipartite graphs.  This raises the intriguing question of giving an exact characterization of the relation, which would be an avenue for further exploration.

On bipartite graphs of diameter $2$, the relation $\requiv{\Tt}$ is the same as isomorphsm.  This is because the only bipartite graphs of diameter $2$ are the bicliques $K_{m,n}$.  Each such biclique is completely characterized by its size parameter $(m,n)$ and so by the relation $\requiv{\Tt}$ by virtue of Theorem~\ref{thm:diam2}.  Thus, our construction of a counter-example of diameter $3$ is optimal in terms of diameter.

Moreover, for graphs that are not necessarily connected, Theorem~\ref{thm:T2} gives a complete characterization of the equivalence relation $\requiv{\Tt_2}$.  It would be interesting to work out similar characterizations of the relations $\requiv{\Tt_d}$ for higher values of $d$.  The remark at the end of Section~\ref{sec:diam3} shows that the equivalence relation $\nse$ is not an exact characterization of $\requiv{\Tt}$, and a complete such characterization remains open.  It might also be possible to simplify the statement of Theorem~\ref{thm:T2}.  In the present version, we require $G$ and $H$ to have the same multiset of \emph{partial sums} of their difference sets.  Is this the same as saying that they have the same difference sets?  This is a purely combinatorial question about numbers: given two sequences $\vec{d}_1$ and $\vec{d}_2$ of non-negative  integers of the same length, if $\psum(\vec{d}_1) = \psum(\vec{d}_2)$, does it follow that $d_1$ and $d_2$ are the same as multisets?

\acknowledgements
In arriving at these results and preparing this manuscript, I have benefitted from very productive discussions with Phokion Kolaitis, Yo\`{a}v Montacute and Wei-Lin Wu.  I am grateful to them.  My search for a counterexample was aided by a computer-based search (though the proof ultimately does not involve computational elements) and I am indebted to my daughter Tarika for inspiring me to take up writing Python programs for this purpose.  Finally, I am grateful to Dan Kr\'{a}l and his collaborators for showing me the construction reported in the remark at the end of Section~\ref{sec:diam3}.

\bibliographystyle{abbrvnat}
\bibliography{refs.bib}

\end{document}